# Human-organ-scale x-ray fluorescence ghost imaging for radioisotope-free diagnostics

E. Levinson[1], R. H. Shukrun[1], N. Viganò[2], and S. Shwartz[1*]

**Affiliations:**

[1] Physics Department and Institute of Nanotechnology and Advanced Materials, Bar-Ilan University, Ramat Gan, 52900 Israel.

[2] Université Grenoble Alpes, CEA, IRIG-MEM, Grenoble, 38000, France

*Corresponding author. Email: Sharon.shwartz@biu.ac.il

**Abstract:** A wide range of diagnostic information in medicine is currently obtained using radioactive tracers. While central to nuclear medicine, these methods are inherently constrained: radiation dose limits repeat examinations, short tracer half-lives and complex logistics restrict access and raise costs, and their relatively poor spatial resolution often necessitates complementary CT or MRI. Here we present a first proof-of-concept demonstration of a non-radioactive alternative based on x-ray fluorescence (XRF) computational ghost imaging (CGI) at the human-organ scale. Using a thyroid phantom filled with iodine solution as a model system, we show that structured illuminations combined with fluorescence detection reconstruct the iodine distribution with high fidelity. This approach eliminates the need for radioactive tracers while preserving image quality, and in principle can reach spatial resolution comparable to CT. Beyond this demonstration, XRF-CGI establishes a generalizable framework for non-radioactive tracer imaging, opening a route toward safer, repeatable, and more accessible diagnostics.

## Introduction

In nuclear medicine, methods such as scintigraphy, positron emission tomography (PET), and single-photon emission computed tomography (SPECT) enable visualization of metabolic processes and tissue-specific uptake using radioactive tracers. In scintigraphy, a detector records gamma emissions from radionuclides to produce 2D functional images. SPECT extends this concept by rotating the detector to reconstruct 3D tracer distributions. PET detects coincident annihilation photons from positron-emitting isotopes, mainly $^{18}$F [1], allowing highly sensitive and quantitatively accurate tomographic imaging of metabolic activity. Together, these methods have become indispensable in modern diagnostics, particularly in oncology and cardiology, where they provide quantitative insights that guide both diagnosis and therapy [2-6].

Despite their wide utility, all nuclear imaging methods fundamentally rely on the radioactive decay of unstable isotopes, whether through direct gamma emission or positron annihilation, imposing unavoidable physical constraints. Ionizing radiation increases the risk of cancer and other diseases, limiting the permissible dose and consequently the frequency of repeated examinations [7,8]. Typically, short-lived radiotracers with half-lives ranging from a few hours to several days are employed, creating major logistical challenges in isotope production, transport, and storage [9,10]. Moreover, the relatively low spatial resolution of SPECT (~8-10 mm) and PET (~4-5 mm under clinical conditions) often necessitates the use of hybrid imaging with CT or MRI, increasing complexity and cost [10,11]. Although advanced reconstruction methods mitigate some of these issues [12-15], they cannot eliminate the intrinsic drawbacks of radioactive imaging itself.

These limitations motivate the search for alternative approaches that avoid radioactive tracers, while maintaining standard spatial resolution and clinically acceptable radiation doses. One intriguing direction lies in x-ray fluorescence (XRF)-based modalities. In XRF, incident x-rays excite inner-shell electrons in the sample, and the subsequent emission of element-specific fluorescent photons provides a direct fingerprint of elemental composition. XRF therefore enables elemental concentration mapping, in close analogy to how SPECT maps radionuclide distributions, and can in principle deliver comparable functional information without reliance on radioactive decays.

A key advantage is that, unlike radioactive tracers, XRF does not rely on ongoing decay: ionizing radiation is present only during x-ray excitation and vanishes immediately once the illumination is turned off. This property has established XRF as a powerful analytical tool across disciplines, including biomedical research and preclinical imaging [16], archaeological studies of pigments and artifacts [17], and environmental monitoring of trace elements in soils and sediment [18]. Early attempts to exploit XRF for medical imaging date back several decades, including fluorescent thyroid scanning in humans, which demonstrated that it could provide sufficient elemental contrast at low dose without radioactive tracers [19]. However, the same studies concluded that conventional XRF implementations are not suitable for human-scale imaging, primarily because they rely on raster scanning with a narrow pencil beam. While pencil beams can produce high-resolution elemental maps, acquisition times scale prohibitively with field of view, making human-organ imaging impractical [20]. As a result, XRF methods have largely remained confined to small-animal studies, and extending it to human-organ scales poses formidable challenges [21]. This bottleneck is not algorithmic but architectural: it arises from the inherently serial nature of pencil-beam XRF, motivating approaches that convert XRF from a point-by-point measurement into a parallel, information-efficient acquisition.

To overcome this challenge, here we introduce a computational approach that replaces sequential pencil-beam scanning with parallel structured illumination and algorithmic reconstruction. In this x-ray fluorescence computational ghost imaging (XRF-CGI) scheme, the element map is recovered by correlating the known illumination patterns with the total fluorescence measured for each pattern [22]. This combines the elemental specificity of XRF with the speed and scalability of computational imaging, enabling rapid acquisition at resolutions that can rival or even surpass CT [23].

In our proof-of-concept experiment, we image a human thyroid phantom and reconstruct the iodine distribution with a spatial resolution of ~4 mm. Although demonstrated here for structural mapping, the same principle could naturally extend to functional imaging using element-targeted contrast agents. Together, these results establish a pathway toward XRF-based medical imaging that combines high spatial precision, rapid acquisition, and the safety of non-radioactive tracers.

## Results

Our approach is based on computational ghost imaging (CGI) [24], in which spatial information is recovered from correlations between a set of known structured illumination patterns and a single integrated (bucket) signal recorded for each pattern. In a discrete representation, the measurement process can be described as [25]:

$$\mathbf{y} = \mathbf{W}\mathbf{x} + \epsilon, \quad (1)$$

where y is the vector of measured bucket fluorescence signals, W contains the applied patterns, x is the unknown elemental distribution, $\epsilon$ is random uncorrelated noise. In practice, W is a matrix and each of its rows is one of the illumination beam structures. We implement CGI here in an XRF setting by projecting a sequence of structured x-ray patterns onto the thyroid phantom and recording, for each pattern, the integrated iodine fluorescence counts in the selected energy window. The elemental map x is reconstructed by solving the inverse problem using compressive sensing or machine-learning methods, both of which substantially reduce the number of required measurements [26,27]. This approach enables large-area elemental imaging without pencil-beam raster scanning.

A schematic of the experimental setup is shown in Fig. 1. A binary amplitude mask mounted on a translation stage is positioned upstream of the object. For each realization, a different region of the mask is illuminated, generating a sequence of structured illumination patterns at the sample plane. Fluorescence photons are detected by a silicon-drift-detector (SDD), which is mounted approximately 45° relative to the incident beam.

The mask consists of a 1.5 mm-thick Ag plate patterned with 0.5x0.5 mm² openings over a 16x16 cm² area, fabricated using high-precision 3D printing [23,28] (open fraction ~ 50%). The object is an anthropomorphic phantom (RSD Inc.) containing a thyroid insert filled with a 200 mg/mL solution of sodium diatrizoate hydrate. The insert area is 7.5x6 cm², representing a human-scale organ. The XRF-CGI reconstruction of the image was performed using the Noise2Ghost (N2G) algorithm, as detailed in the Methods.

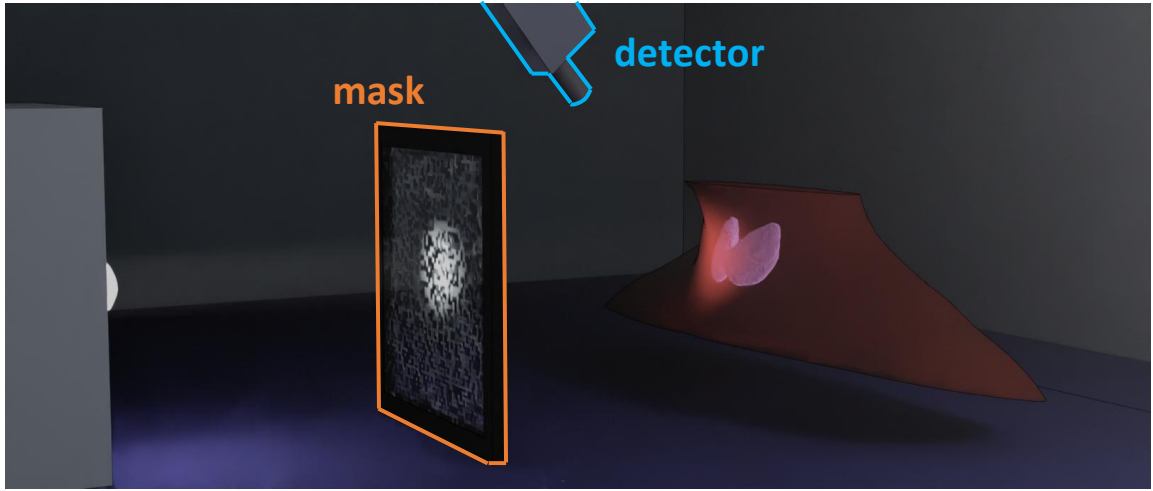

**Fig. 1.** A schematic of the experimental system. Incident x-ray photons pass through the mask and excite the iodine solution within the thyroid phantom, generating characteristic fluorescence emission. The emitted fluorescence photons are collected by an energy-resolving detector, which discriminates the iodine fluorescence lines from the scattered background.

To demonstrate human-scale XRF-CGI, Fig. 2a shows the reconstruction of the thyroid insert filled with an iodine solution, embedded within the anthropomorphic phantom. For comparison and validation, Figs. 2b and 2c show the corresponding flat-panel detector (FPD) transmission images of the isolated thyroid insert and of the entire phantom, respectively.

The XRF-CGI reconstruction accurately recovers the thyroid geometry and internal structure and is consistent with the transmission reference. Notably, the thyroid is more clearly delineated in the fluorescence reconstruction than in the transmission image of the full phantom. This difference reflects the intrinsic selectivity of XRF: the fluorescence signal originates primarily from iodine-containing regions, whereas transmission imaging integrates attenuation from all tissues along the beam path. As a result, XRF-CGI naturally suppresses anatomical background and isolates the elemental distribution of interest.

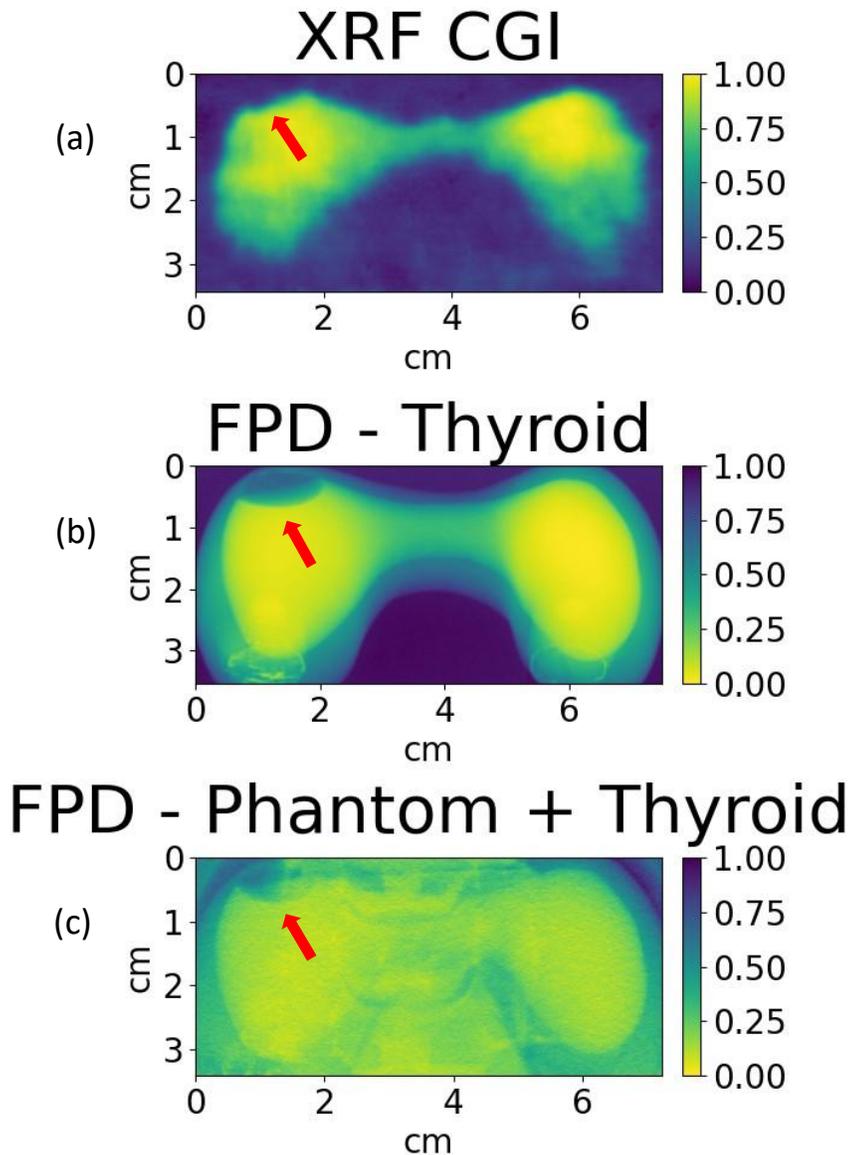

**Fig. 2.** Images of the thyroid phantom filled with the iodine solution. (a) XRF-CGI reconstruction obtained using the self-supervised N2G approach with 5000 illumination realizations, demonstrating faithful recovery of the thyroid insert structure (b) Transmission Flat-panel detector (FPD) image of the thyroid insert measured separately from the phantom. (c) FPD transmission image of the complete anthropomorphic phantom. All images are displayed using normalized intensity scales. For the XRF images, the color bar corresponds to the fluorescence yield, while for the transmission images it corresponds to the inverse of the transmission.

In the upper left region of the thyroid (indicated by arrows in all panels of Fig. 2), a small missing section is visible, especially in the isolated insert and in our XRF-CGI reconstruction. This feature originates from an air bubble trapped during filling of the cavity with the iodine solution. Resolving this localized void highlights the ability of XRF-CGI to capture internal inhomogeneities within a human-scale organ phantom.

Having demonstrated the capability of our method to image the thyroid insert, we next assess its accuracy. Fig. 3a shows one-dimensional profiles through the center of the thyroid insert, extracted from the FPD reference image (Fig. 2b) and from the N2G reconstruction. The two profiles exhibit nearly identical shapes of the thyroid lobes, confirming that our approach accurately recovers the iodine distribution. Agreement is obtained both on the organ scale and for internal features: the separation between the two thyroid lobes is $45.2 \pm 0.2$ mm in the reference image and $44.7 \pm 4.5$ mm in the reconstruction. The air bubble void has a horizontal width of $9.0 \pm 0.2$ mm in the reference and $8.2 \pm 4.5$ mm in the reconstruction. Small differences are expected because XRF-CGI is selective to iodine-bearing regions, whereas transmission integrates attenuation over the entire thyroid volume, including surrounding tissue.

To provide an independent and quantitative estimate of spatial resolution, Fig. 3b presents the Fourier ring correlation (FRC) curve computed between two N2G reconstructions, each obtained from 3000 independent illumination realizations. Using the standard FRC threshold criterion (see Methods) we obtain an effective spatial resolution of approximately 4.5 mm. This value is consistent with the smallest resolved features in Fig. 2 and with the profile comparison in Fig. 3a, confirming that the reconstruction retains millimeter-scale structural information across the human-scale field of view.

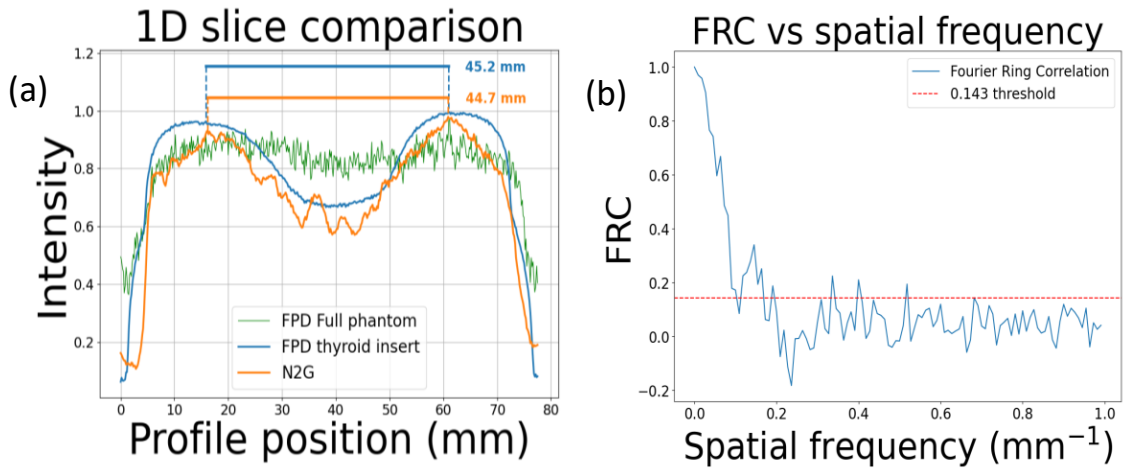

**Fig. 3.** (a) One-dimensional profiles through the center of the thyroid. Shown are lines extracted from the FPD image of the complete phantom, the FPD image of the thyroid insert separated from the phantom, and from a N2G reconstruction. The profiles of the isolated thyroid and the N2G reconstruction exhibit closely matching shapes and similar widths, with lobe separations of 45.2 mm and 44.7 mm, respectively, demonstrating strong agreement between the fluorescence-based reconstruction and the reference transmission measurement. (b) Fourier ring correlation (FRC) between two independent ghost-imaging reconstructions (solid line). The dashed red line indicates the 0.143 threshold. The first crossing occurs at 0.109 mm$^{-1}$, giving a spatial resolution of approximately 4.5 mm.

Having established that high-quality images can be demonstrated under these conditions, we next quantify how image quality scales with the number of mask realizations, which directly determines the ionizing-radiation dose in XRF-CGI. In Fig. 4a,b we show the dependence of the Contrast-to-Noise Ratio (CNR) and Mean-squared-error (MSE) on the number of realizations, for data without binning and with a binning ×16 factor. Binning was introduced to accelerate post-acquisition processing and to assess how much spatial information could be discarded before the reconstruction quality noticeably degraded. As seen in Fig. 4a, both cases follow the same overall trend, although the binned data yield higher CNR at low realization counts. In Fig. 4b, the binned data exhibit a somewhat faster decrease in the initial stages. Beyond this initial drop, however, the curves converge to similar values, indicating that even with substantial binning the reconstruction quality was not compromised in this experiment.

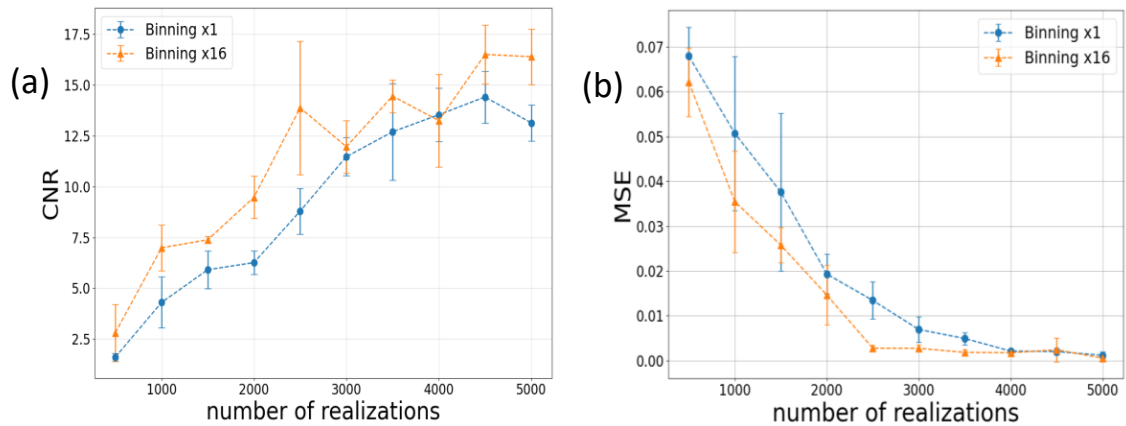

**Fig. 4.** Dependence of (a) Contrast-to-Noise Ratio (CNR) and (b) Mean-squared-error (MSE) on the number of realizations. Each data point represents the mean of five independent reconstructions for a given number of realizations, and different colors denote distinct binning factors. Lines are guides to the eye. Error bars indicate the standard deviation.

To further illustrate the dependence of the image quality on the number of realizations, Fig. 5 shows reconstructions obtained with increasing numbers of illumination realizations. Image quality improves monotonically as additional realizations are acquired, with thyroid boundaries and internal features becoming progressively clearer. These examples visually corroborate the CNR and MSE trends and highlight the acquisition-quality tradeoff that sets the dose requirements of XRF-CGI.

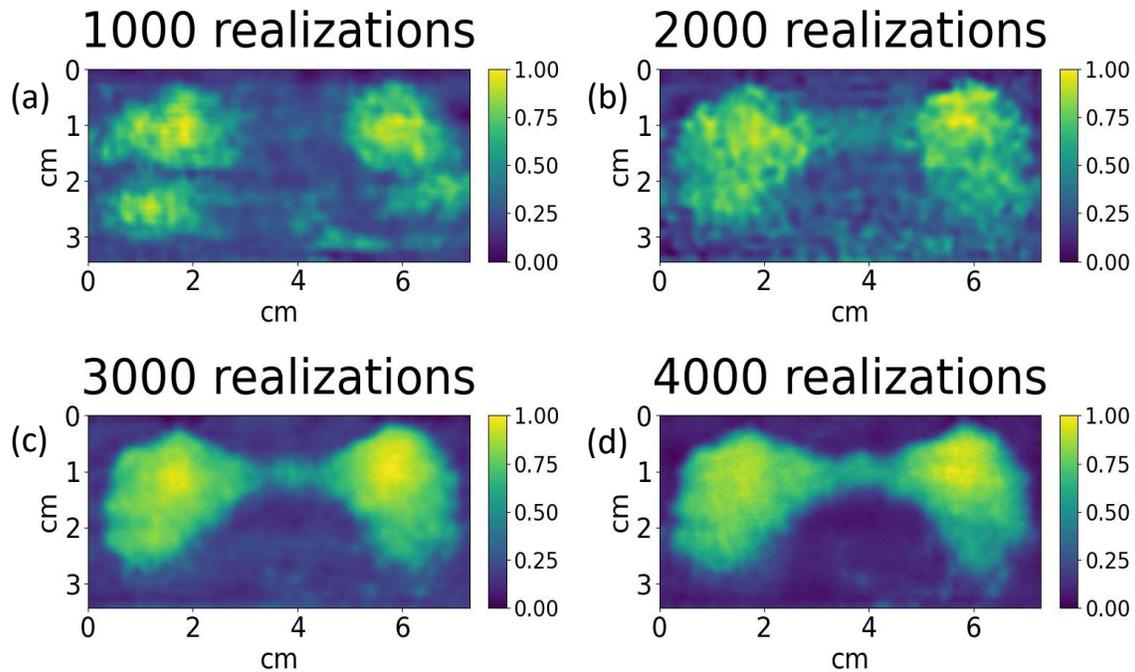

**Fig. 5.** Dependence of the N2G reconstructions on the number of realizations. Reconstructions of the thyroid phantom are shown for (a) 1,000, (b) 2,000, (c) 3,000, and (d) 4,000 realizations. The two thyroid lobes are discernible even at low realization counts, and image quality improves progressively with additional realizations. All images are displayed using normalized intensity scales.

**Discussion**

Our human-organ-scale demonstration of XRF-CGI with a thyroid phantom establishes a proof-of-concept for retrieving the spatial distribution of non-radioactive markers in clinically sized organs. A particularly clear indicator of reconstruction fidelity is the visibility of the air bubble within the thyroid insert: despite the non-ideal, early-stage implementation, the reconstruction resolves a local inhomogeneity in the marker distribution. This sensitivity suggests that XRF-CGI can capture clinically relevant structural variations at human scale.

Because non-radioactive elemental markers can, in many cases, be chemically analogous to their radioactive counterparts, XRF-CGI has the potential to provide functional information similar to PET or SPECT while eliminating the logistical burden and radiation risks associated with radioisotopes. Moreover, XRF-CGI [22] and x-ray ghost imaging [23] tomography have been shown to produce images with spatial resolution approaching that of CT. This could reduce reliance on CT co-registration, which is routinely used to compensate for the limited spatial resolution of PET/SPECT, thereby lowering overall patient dose.

Translating the present proof-of-concept into a clinically practical modality requires substantial reductions in both acquisition time and dose. We therefore estimate the measurement time and dose based on our current parameters and on realistic upgrades to the hardware and reconstruction pipeline. In the present implementation we acquired 5000 realizations with 20 s per realization, giving a total acquisition time of nearly 30 h, clearly beyond any clinical workflow. This long duration is primarily dictated by limited fluorescence collection efficiency: the SDD used here has an efficiency of ~13% at the iodine Kα fluorescence line (~28 keV) and an active area of 17 mm$^2$. Replacing it with modern CdTe detector arrays, comprising 20 elements, with near unity efficiency at 28 keV and an effective area of 340 mm$^2$, would increase the collected fluorescence yield by a factor of ~150, reducing acquisition time proportionally. Additional reductions could be obtained by decreasing the detector-phantom distance, which increases the detected rate approximately with the inverse square of the distance. However, operating at shorter standoff requires careful optimization to control scattered-radiation backgrounds (especially in anatomically heterogeneous settings) and to mitigate dead time through appropriate choices of shaping time and front-end electronics.

A second bottleneck is spectral mismatch between the excitation spectrum and the iodine absorption edge. Our current tube spectrum (80 kVp) is not optimized for iodine excitation. We estimate that only 5% of the source spectrum contributes effectively, and geometric constraints applied to reduce background further reduce the flux that reaches the target thyroid. By shifting to an excitation spectrum centered around 40 keV and using filtration that suppresses non-contributing photons while transmitting those above the iodine K-edge [16], the fraction of photons effectively driving fluorescence can increase by nearly two orders of magnitude with present-day technology. Finally, continued progress in reconstruction, particularly via advanced machine-learning or compressive strategies, should further decrease the number of realizations needed for high-fidelity reconstructions by an additional order of magnitude.

Taken together, these improvements indicate a realistic pathway toward clinically relevant acquisition times on the order of seconds, comparable or potentially faster than current nuclear-medicine protocols, while using radioisotope-free tracers. Looking further ahead, quasi-monochromatic sources such as inverse-Compton systems could further increase the spectral efficiency and reduce dose [29], potentially pushing acquisition times below today's state-of-the-art functional imaging while preserving the key safety advantage of eliminating radioactive decay.

Finally, we emphasize that XRF-CGI is not specific to the thyroid or to iodine. In principle, it can be generalized to other organs and to a broad class of non-radioactive markers, including elements for which no suitable clinical radioisotope exists today.

## Methods

The detector used for fluorescence measurements was an Amptek FAST silicon-drift-detector (SDD), positioned approximately 14 cm from the phantom to balance primary fluorescence signal and scattered background. The x-ray source was operated at 80 kVp and 2 mA.

Spatial binning was applied only to the structured illumination patterns, while the measured bucket fluorescence signals were kept unchanged. This was done by averaging non-overlapping N×N pixel blocks over each pattern.

The reconstructions presented were obtained using Noise2Ghost (N2G), a self-supervised convolutional neural network (CNN) method for CGI [25]. N2G partitions the realizations into subsets to form sub-reconstructions and trains a CNN to transform those sub-reconstructions such that they match independent measurements in the forward model. We consider the forward model we introduced in Eq. (1). N2G seeks a neural network parametrization $\hat{\theta}$ that minimizes the following expression:

$$\hat{\theta} = \underset{\theta}{\operatorname{argmin}} \; \frac{1}{2} \sum_k \sum_{i \neq k} ||W_i N_\theta(x_k) - y_i||_2^2 + \lambda R(N_\theta(x_k)), \qquad (2)$$

where the indices i and k refer to different measurement subsets and related sub-reconstructions, and K is the number of subsets. The expected reconstruction error will then be:

$$E_{b, i \neq k} ||W_i N_\theta(x_k) - b_i||_2^2 + E_{\epsilon, i \neq k} ||\epsilon_i||_2^2, \qquad (3)$$

where the first term is the supervised reconstruction error and does not depend on the noise. The final reconstruction is obtained by averaging the network outputs for all subsets:

$$\hat{x} = \frac{1}{K} \sum_k N_{\hat{\theta}}(x_k). \qquad (4)$$

To quantify reconstruction quality, we used MSE, CNR, and FRC.

For the MSE, the deviations between corresponding pixels in the reconstructed and reference images are computed as:

$$MSE = \frac{1}{N} \sum_{j=1}^{N} (I_j - R_j)^2, \qquad (5)$$

where $I_j$ and $R_j$ denote the intensity values of the reconstructed and reference image pixels, respectively, and $N$ is the number of pixels. Because an external ground-truth image is unavailable, the reference image is defined as the average of five independent N2G reconstructions obtained using 5000 realizations. The MSE therefore measures convergence toward this stable reference.

To quantify contrast, we computed the CNR, which accounts for the signal and background levels as well as signal variability:

$$CNR = \frac{|S_A - S_B|}{\sigma_A}, \quad (6)$$

where $S_A$ and $S_B$ are the mean intensities in the signal and background regions, respectively, and $\sigma_A$ is the standard deviation within the signal region. We normalize by $\sigma_A$ rather than the background standard deviation because the background level decreases strongly with the number of realizations, which would otherwise artificially inflate the CNR at large realization numbers.

FRC is used as an independent and objective estimate of spatial resolution. Two reconstructions obtained from statistically independent sets of realizations are Fourier-transformed, and their normalized cross-correlation is computed over concentric rings in Fourier space as a function of spatial frequency r:

$$FRC(r) = \frac{\sum_{i \in r} F_1(r_i) F_2(r_i)^*}{\sqrt{\sum_{i \in r} |F_1(r_i)|^2 \sum_{i \in r} |F_2(r_i)|^2}}, \quad (7)$$

where $F_1$ and $F_2$ are the Fourier transforms of the two reconstructions, and the summation is performed over all Fourier pixels $r_i$ on a ring of radius r. The resolution is defined as the spatial frequency at which the FRC curve crossed the 0.143 threshold, a commonly used empirical criterion [30].

**Acknowledgements**

This work is supported by the Israel Science Foundation (ISF) (847/21)

**Author contributions**

E.L. performed all experiments and led the data analysis. R.H.S. conceived the project and provided feedback on the manuscript. N.V. developed the Noise2Ghost algorithm used for the image reconstructions. S. Shwartz supervised the research. E.L. and S. Shwartz wrote the manuscript. All authors contributed to the work presented here and to the final paper.

# Human-organ-scale X-ray fluorescence ghost imaging for radioisotope-free diagnostics

E. Levinson[1], R. H. Shukrun[1], N. Viganò[2], and S. Shwartz[1*]

# Supplementary Information

**Supplementary note 1. N2G code parameters**

Reconstructions were performed using the N2G method, which includes several tunable parameters that affect reconstruction fidelity. Supplementary Table 1 lists the parameter values used for the reconstructions shown in the main text.

| Parameter | Binning x1 | Binning x16 |
| --- | --- | --- |
| Num. of features | 20 | 20 |
| Num. of levels | 3 | 3 |
| Regularization value | $2 \times 10^{-6}$ | $2 \times 10^{-6}$ |
| Num. of permutations | 3 | 5 |
| Num. of splits | 4 | 5 |
| Epochs | 3072 | 3072 |
| Learning rate | 0.001 | 0.001 |

**Supplementary Table 1**. N2G parameters used for the reconstructions.

**Supplementary note 2. Phantom orientation**

The anthropomorphic phantom was positioned with an upward tilt of approximately 45° relative to the incident x-ray beam. This orientation was chosen so the thyroid insert would face both the excitation beam and the SDD, improving fluorescence collection efficiency by reducing self-absorption along the detection path.

**Supplementary note 3. Binning**

Additional reconstructions were performed using multiple binning factors to assess degradation in reconstruction quality with increasing loss of spatial information. Fig. S1a, b show the corresponding CNR and MSE, respectively. Both metrics deteriorate for the largest binning factor, indicating loss of spatial detail beyond this level of binning.

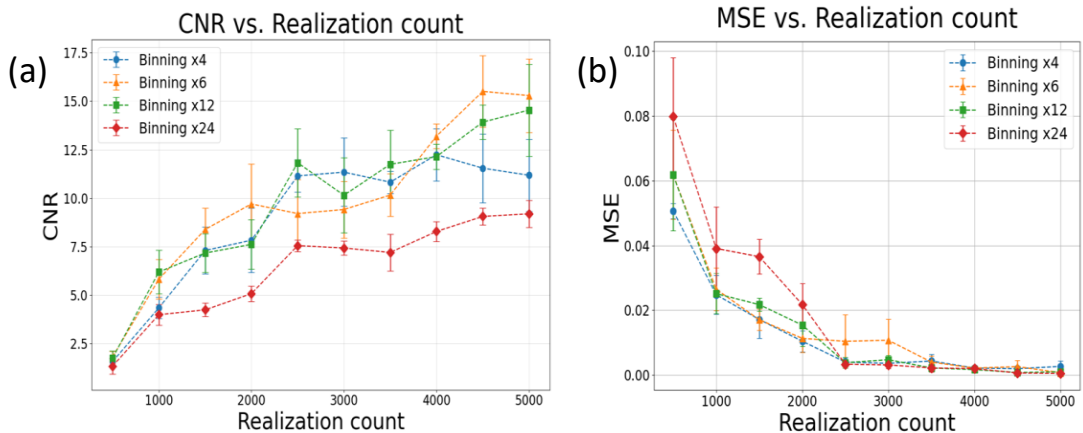

**Fig. S1.** (a) CNR. (b) MSE. Each data point represents the mean of five independent reconstructions for a given number of realizations, and different colours denote distinct binning factors. Lines are guides to the eye. Error bars indicate the standard deviation.

**Supplementary note 4. Energy window selection**

The reconstruction energy window was set approximately ±1 KeV around the Iodine $K_\alpha$ peak. Including the $K_\beta$ region was tested but increased additional noise and was therefore omitted. The measured spectrum and the selected energy window are shown in Fig. S2.

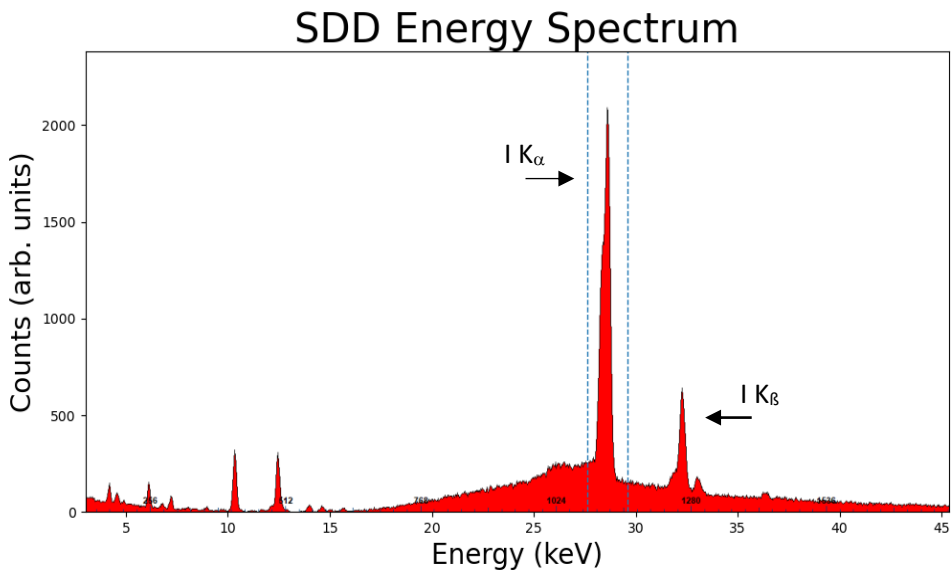

**Fig. S2.** The SDD energy spectrum. The reconstruction energy window (dashed lines) is indicated. The iodine $K_\alpha$ and $K_\beta$ fluorescence lines at 28.6 KeV and 32.3 KeV, respectively, are visible. Weaker Pb L-shell fluorescence peaks at lower energies are also observed, originating from surrounding shielding.